\documentclass{aa}
\usepackage{graphicx}
\usepackage{txfonts}
\usepackage{amsmath,amssymb}
\usepackage{mathrsfs}	
\usepackage{mathtools}  
\usepackage{color}
\usepackage{colortbl}
\usepackage{array}

\usepackage[bookmarks=true]{hyperref}
\hypersetup{
  colorlinks=true,
  pdfborder=2 2 0.,
  linkcolor=red,
  anchorcolor=black,
  citecolor=blue,
  urlcolor=black,
}

\usepackage{natbib}
\bibpunct{(}{)}{;}{a}{}{,} 

\newcommand{\Mstar}{M_\star}
\newcommand{\mstar}{M_\star}
\newcommand{\MHI}{M_{\rm HI}}
\newcommand{\fj}{f_j}
\newcommand{\fM}{f_M}

\newcommand{\fr}{f_R}
\newcommand{\fb}{f_{\rm b}}
\newcommand{\Omegab}{\Omega_{\rm b}}
\newcommand{\Omegam}{\Omega_{\rm m}}

\newcommand{\Mh}{M_{\rm h}}
\newcommand{\jh}{j_{\rm h}}
\newcommand{\Rh}{R_{\rm h}}
\newcommand{\Vh}{V_{\rm h}}
\newcommand{\sigl}{\sigma_\lambda}
\newcommand{\sigc}{\sigma_c}
\newcommand{\sigm}{\sigma_{M_\star-M_{\rm h}}}
\newcommand{\rlc}{r_{\lambda-c}}
\newcommand{\Rd}{R_{\rm d}}
\newcommand{\Vc}{V_{\rm c}}
\newcommand{\Vdm}{V_{\rm DM}}
\newcommand{\Vstar}{V_\star}
\newcommand{\jstar}{j_\star}
\newcommand{\RTF}{R_{\rm TF}}

\newcommand{\vf}{V_{\rm flat}}
\newcommand{\vtwo}{V_{2.2}}

\defcitealias{PFM19}{PFM19}
\defcitealias{MMW98}{MMW98}

\begin{document}

\title{The impact of the halo spin-concentration relation on disc scaling laws}
\titlerunning{Impact of halo spin-concentration relation on scaling laws}
\authorrunning{L. Posti et al.}

  \author{Lorenzo Posti\inst{1}\fnmsep\thanks{lorenzo.posti@astro.unistra.fr},
          Benoit Famaey\inst{1},
          Gabriele Pezzulli\inst{2,3},
          Filippo Fraternali\inst{3},
          Rodrigo Ibata\inst{1},
          Antonino Marasco\inst{4}
          }
  \institute{Universit\'e de Strasbourg, CNRS UMR 7550, Observatoire astronomique de Strasbourg, 
             11 rue de l'Universit\'e, 67000 Strasbourg, France.
        \and
        Department of Physics, ETH Zurich, Wolfgang-Pauli-Strasse 27, 8093 Zurich, Switzerland
         \and
             Kapteyn Astronomical Institute, University of Groningen,
  	    	  P.O. Box 800, 9700 AV Groningen, the Netherlands
         \and
             INAF - Osservatorio Astrofisico di Arcetri, Largo E. Fermi 5, 50127, Firenze, Italy
             }
      \date{Received XXX; accepted YYY}

  \abstract{Galaxy scaling laws, such as the Tully-Fisher, mass-size and Fall relations, can
  provide extremely useful clues on our understanding of galaxy formation in a cosmological context. 
  Some of these relations are extremely tight and well described by one single parameter (mass),
  despite the theoretical existence of secondary parameters such as spin and concentration,
  which are believed to impact these relations.
  In fact, the residuals of these scaling laws appear to be almost uncorrelated with each other,
  posing significant constraints on models where secondary parameters play an important role.
  Here, we show that a possible solution is that such secondary parameters
  are correlated amongst themselves, in a way that removes correlations in observable space.
  In particular, we focus on how the existence of an anti-correlation between the dark
  matter halo spin and its concentration -- which is still debated in simulations -- can
  weaken the correlation of the residuals of the Tully-Fisher and mass-size relations.
  Interestingly, using simple analytic galaxy formation models, we find that this happens only
  for a relatively small portion of the parameter space that we explored, which suggests that
  this idea could be used to derive constraints to galaxy formation models that are still
  unexplored.
  }
  \maketitle

\section{Introduction} \label{sec:intro}

The fact that some of the most basic and fundamental dynamical properties of disc galaxies,
such as mass, velocity and angular momentum, are very simply correlated to one another
is a crucial testimony of how galaxies assembled in our Universe. The relationships
between such structural and dynamical properties, often called scaling laws, are invaluable 
probes of how galaxies have formed and evolved \citep{McGaugh+00,Dutton+07,Lelli+16a,
Posti+19b}.

The simple power-law shapes of many observed scaling relations are commonly used as a
test-bed for theoretical galaxy formation models. The observed slopes and normalisations
of e.g. the mass-velocity relation \citep[][hereafter TF relation]
{TullyFisher77, McGaugh+00}, the mass-size relation \citep[][hereafter MS relation]{Kormendy77},
the mass-angular momentum relation \citep[][hereafter Fall relation]{Fall83} can in
principle directly constrain the galaxy -- halo connection,
which is the backbone of any galaxy formation model in the $\Lambda$ Cold Dark Matter
($\Lambda$CDM) cosmogony \citep{MMW98,Lapi+18,Posti+19b}. The assembly and the structure
of CDM halos is well understood and we know that they are fully rescalable, i.e. there
exist simple power-law scalings between mass, velocity, angular momentum, size etc.
\citep[e.g.][]{MvdBW10}. These relations for halos immediately translate into those
for galaxies through some fundamental parameters of the galaxy -- halo connection such
as the efficiency at turning baryons into stars or the efficiency at retaining the
angular momentum initially acquired from the gravitational torques exerted by
nearby structures \citep[e.g.][]{Posti+19b}.

In $\Lambda$CDM the TF relation is set to first order by the stellar-to-halo mass
relation \citep[e.g.][]{NavarroSteinmetz00}. Halos acquire angular momentum through
tidal torques at turnaround \citep[e.g.][]{Peebles69} and when the galaxy disc
settles in the centre, incorporating a given fraction of that angular momentum,
its size will then depend on the amount of angular momentum of the halo
\citep[e.g.][]{FallEfstathiou80}. The fact that the baryonic TF relation
\citep{McGaugh+00,Lelli+16a}, relating the total mass in stars and cold gas to
the flat circular velocity, appears tighter than the stellar TF relation makes the
picture more complicated, as it indicates that the scatter in the stellar-to-halo
mass relation might be related to the scatter in cold gas mass. As such, the small
scatter of the baryonic TF is still challenging for our current understanding
of galaxy formation \citep{diCintioLelli17,Desmond17}. 

The residuals of the TF around the mean also carry important
information that are sensitive to the details of the galaxy formation
process \citep[e.g.][]{CourteauRix99,Pizagno+07,vanderKruitFreeman11}. In
particular, considering for example stellar mass and rotational velocity as two
fundamental properties of a galaxy, if the residuals of the TF were found to correlate
with a third property (e.g. galaxy size) it would then mean that the TF is not a
fundamental law, but just a projection of a more general $M-V-R$ relation. 
Thus, many have looked for additional quantities that correlate with the TF residuals,
only to find no significant correlations 
\citep[e.g.][]{Barton+01,Kannappan+02,McGaugh05,Courteau+07}. 
In particular, the fact that the TF residuals do not appear to correlate with the disc
size \citep[][{but see also \citealt{ManceraPina+20} who instead find a correlation in
the dwarf galaxy regime}]{Lelli+16a,Ponomareva+18} nor with the residuals of the MS relation
\citep{McGaugh05,Desmond17} poses several challenges to our understanding of disc galaxies. 
For instance, \citet[][see also \citealt{DuttonvdBosch12}]{Dutton+07} generated rather
sophisticated semi-empirical models, based on the assumption that the angular momentum of
the galaxy is proportional to that of the halo, and found it complicated to find a model
that matched the observed scaling laws while having negligible correlation in the TF
residuals versus MS residuals (when calculated at a fixed mass, not luminosity, see e.g.
Fig. 10 in \citealt{Dutton+07}).
In fact, this issue has later been used to argue that the observed absence of correlations
in the TF and MS residuals provides evidence against the hypothesis that the galaxy's and
halo's specific angular momenta are directly proportional, leaning towards an empirical,
but less physically motivated, anti-correlation between galaxy size and halo concentration
\citep{Desmond+19,Lelli+19}.

However, these simple inferences often neglect the existence of correlations between the
parameters of the theory themselves. For instance, it has been proposed that the halo spin
and concentration are in fact correlated with each other \citep{Maccio+07,Johnson+19}, and
furthermore, there are reasons to expect that the stellar-to-halo mass fraction and angular
momentum fraction are also correlated \citep[e.g.][]{DuttonvdBosch12}.
In this paper, we examine the question of how do these correlations impact our expectations
on the residuals of galaxy scaling relations. We focus in particular on the impact of two
physical effects on the residuals of the TF, MS and Fall relation: we allow (i) the halo
spin to be anti-correlated with the halo concentration \citep[as it is observed in N-body
simulations, e.g.][]{Maccio+07} and (ii) the stellar-to-halo mass fraction to be correlated
with the stellar-to-halo specific angular momentum fraction \citep[as it is expected if the
formation of disc galaxies proceeds inside-out, e.g.][]{RF12,Pezzulli+15,Posti+18a}. 
Our goal here is then to understand the effect of these two ingredients in a rather
isolated and simplified context. Thus, we generate semi-empirical models based on the
assumption that the galaxy's and halo's angular momenta are related, but which we keep
deliberately simple in order to answer the question of whether the addition of the two new
ingredients mentioned above can help in reproducing the observed disc scaling laws.

We note, however, that the correlation of the scaling laws residuals are intrinsically
noisy observables that provide typically very poor statistical inference compared to,
for instance, the global shape (slope, normalisation) of the scaling laws themselves.
In fact, (i) just by definition they rely on a fit of the observed scaling law,
which is itself subject to systematic uncertainties; (ii) estimating a correlation
coefficient from a discrete distribution of points is sensitive to Poisson
noise for the sample sizes typically considered here (hundreds/thousands); 
(iii) covariance/correlation estimators are sensitive to outliers and biases in the
population samples.
To mitigate these limitations we use the SPARC catalogue of nearby spirals \citep{SPARC},
which is of the highest quality for dynamical studies and which has already been used to
study this topic \citep{Desmond+19,Lelli+19}. However, even though this is currently the
best available data-set, it does not remove all the issues mentioned above.

Throughout the paper we use a fixed critical overdensity parameter
$\Delta=200$ to define virial masses, radii etc. of dark matter haloes and
the standard $\Lambda$CDM model, with parameters estimated by the \cite{Planck18}:
baryon fraction $\fb\equiv\Omegab/\Omegam \simeq 0.157$ and Hubble constant
$H_0=67.4$ km s$^{-1}$ Mpc$^{-1}$.

\section{Models} \label{sec:model}

We describe here the ingredients and the procedure that we use to build our analytic
models. These borrow heavily from the seminal paper of \citet[][hereafter MMW98]{MMW98}.
These simple models neglect the contribution of the gas to the dynamics, hence we will
restrict the comparison to stellar-dominated galaxies in the SPARC sample of \citet{SPARC}.

\subsection{Dark Matter halo population} \label{sec:dm}

We generate a population of dark matter halos as follows.

{\bf Mass function.} We start by sampling an analytic halo mass function, which is a
well-known property of the cosmological model we adopt. In particular we use the halo
mass function from \citet[][evaluated and sampled using the code \texttt{hmf},
\citealt{HMF}]{Tinker+08}.

{\bf Spin.} Each halo is assigned a spin parameter $\lambda \equiv \jh/(\sqrt{2} \Rh\Vh)$
\citep[see][]{Bullock+01}, where $\Rh$, $\Vh$ and $\jh$ are the virial radius, velocity
and specific angular momentum respectively. The spin parameter $\lambda$ is drawn from a
log-normal distribution with mean $\log \overline{\lambda} = -1.45$ and scatter
$\sigl=0.22$ dex \citep[e.g.][]{Maccio+07}.

{\bf Halo density profiles.} We assume that each DM halo follows a \citet[][hereafter NFW]{NFW}
profile, which is characterised by 2 parameters: the virial mass $\Mh$ and concentration $c$.
These two follow a well-established anti-correlation, known as the $c-\Mh$ relation,
such that more massive halos are less concentrated. We assign halo concentration following the
parametrisation of the $c-\Mh$ relation from \cite{DuttonMaccio14}, with intrinsic scatter of
$\sigc = 0.11$ dex \citep[but this could be as high as 0.16 dex, see e.g.][]{DiemerKravtsov15}.

{\bf Spin-concentration anti-correlation.} We allow the spin $\lambda$ and the
concentration $c$ of DM halos to be negatively correlated, as found in numerical N-body
simulations \citep[e.g.][]{Maccio+07}. 
The existence of this negative correlation might
be a result of the assembly history of haloes, i.e. haloes that have assembled later spin
faster and have shallower density profiles due to the material deposited in the outskirts
by recent mergers \citep[e.g.][]{Johnson+19}. However, it is still debated whether this
correlation is a robust prediction of $\Lambda$CDM and how much it is sensitive to sample
selection, as including or excluding halos that are defined to be unrelaxed seems to have an
effect on the measured strength of this correlation \citep[e.g.][]{Maccio+07,Neto+2007}.
Since this issue does not appear to be fully settled, it is worthwhile asking what happens
to the predictions of a semi-empirical galaxy formation model that includes this correlation.
Since we could not find any analytic description of this correlation,
we parametrise the $\lambda-c$ correlation with a correlated 2-D normal distribution in
$\log\, \lambda$ and $\log\,c$, with correlation coefficient $\rlc$ that is a free
parameter of the model. One of our main results hereafter will be to show how only a tight
range of values of this parameter allow us to reproduce the absence of correlation of the
residuals of observed scaling relations. In Appendix~\ref{app:bolshoi} we perform a simple
exploration of the public halo catalogues of the dark matter-only Bolshoi simulation
\citep{Klypin+16,Rodriguez-Puebla+16}, where we estimate that the correlation coefficient
is of the order of $\rlc\simeq -0.3$.

\subsection{Galaxies} \label{sec:galaxies}

We assign a single galaxy to each dark matter halo, thus assuming that each
galaxy is central to its halo.

{\bf Stellar mass.} Each halo hosts a galaxy whose stellar mass $\mstar$ follows a given
stellar-to-halo mass relation, in this case not from abundance matching but from
\cite{Posti+19b}. This is an unbroken
power-law relation which is valid for spiral galaxies, and was in fact derived using data
from the SPARC galaxy catalogue. We assume a scatter of
$\sigm=0.15$ dex, similar to what it is typically expected for the $\Mstar-\Mh$ relation
\citep[e.g.][]{Moster+13} and measured using a variety of techniques
\citep[][see also \citealt{WechslerTinker18}, and references therein, for a recent review]
{More+11,Yang+09,ZuMandelbaum15}. 

{\bf Stellar density profiles.} We assume that galaxies are thin exponential discs, with
stellar surface density $\Sigma_\star = \Sigma_0\,\exp(-R/\Rd)$, where $\Rd$ is the
disc scale-length and $\Sigma_0=\Mstar/2\pi\Rd^2$ is the central surface density. As we
neglect the presence of a gas disc, we will restrict the comparison to stellar-dominated
galaxies in the SPARC sample, i.e. $\MHI/\Mstar < 1$.

{\bf Circular velocity.} The circular velocity of our model galaxies is made up from the
contribution of dark matter ($\Vdm$) and stars ($\Vstar$) as $\Vc=\sqrt{\Vdm^2+\Vstar^2}$,
where both $\Vstar$ and $\Vdm$ are analytic functions for an exponential disc and an NFW
profile respectively \citep{Freeman70,NFW}. 

{\bf Disc scale-length.} We calculate the disc scale-lengths using the iterative procedure
proposed by \citetalias{MMW98}. The galaxy disc specific angular momentum, assuming that
stars are on circular orbits, is:
\begin{equation} \label{eq:jstar}
\jstar \equiv \frac{J_\star}{M_\star} = \frac{2\pi}{M_\star} \int {\rm d}R \,R^2 \,
    \Vc\,\Sigma_\star.
\end{equation}
If the rotation curve had been perfectly constant and equal to $\Vh$, e.g. in the case of
a dominant singular isothermal halo \citepalias{MMW98}, then the disc specific angular
momentum would have been equal to $2\Rd\Vh$; thus, for convenience, we introduce the
ratio of $\jstar$ to $2\Rd\Vh$, i.e.\footnote{
Note that the parameter $\xi$ is just the inverse of what \citetalias{MMW98} call $\fr$. We
use a different notation here not to confuse the reader with the ratio of disc size to
halo virial radius that we call $\fr$ in a previous paper \citep{Posti+19b}.
}
\begin{equation} \label{eq:xi}
\xi = \frac{1}{2} \int {\rm d}u \,u^2\, \frac{\Vc(u\Rd)}{\Vh}\,{\rm e}^{-u},
\end{equation}
where $u=R/\Rd$, such that $\jstar = 2\Rd\Vh\xi$. In our model, galaxies acquire angular
momentum from the same tidal torques that set the dark halo spinning, thus we can relate
the stellar angular momentum $\jstar$ to the halo spin parameter $\lambda$ by introducing
the retained fraction of angular momentum $\fj\equiv\jstar/\jh$, from which it follows
that $\jstar = \sqrt{2}\lambda\fj\Rh\Vh $. Rearranging this, together with
Eq.~\eqref{eq:jstar}-\eqref{eq:xi}, we can write the relation between disc size and halo
size as
\begin{equation} \label{eq:Rd}
\Rd = \frac{1}{\sqrt{2}}\lambda\,\fj\,\xi^{-1}\Rh.
\end{equation}
In practice, to solve for $\Rd$ for each model galaxy we have to proceed iteratively (as in
\citetalias{MMW98}). We start with a first guess for $\Rd$ by setting $\xi=1$, which is the
case of an isothermal halo that gives $\jstar=2\Rd\Vh$, and an expression for $\fj$ that is
discussed below. With this guess for $\Rd$ we proceed to compute $\xi$ as in Eq.~\eqref{eq:xi}
and subsequently $\Rd$ again as in Eq.~\eqref{eq:Rd}. 
We iterate this procedure 5 times for each galaxy, which is enough to guarantee convergence
on the value of the disc scale length.

{\bf Retained fraction of angular momentum.} We finally allow the ratio of stellar-to-halo
specific angular momentum $\fj\equiv\jstar/\jh$ to be a function of the stellar mass fraction 
$\fM\equiv\Mstar/\Mh$. This kind of models are commonly known as \emph{biased collapse}
models, where stars are formed from the inside-out cooling of gas, from the angular momentum
poorest material to the angular momentum richest material \citep{DuttonvdBosch12,Kassin+12,
RF12,Posti+18a}.
Thus we have
\begin{equation} \label{eq:fjfm}
    \fj \propto \fM^s,
\end{equation}
where $s$ is a free parameter of the model. We assume an intrinsic scatter of
$\sigma_{\fj}=0.07$ dex on this relation, which is consistent with the analysis of
\cite{Posti+19b} on the local disc scaling laws. We note that the case of specific angular
momentum equality between stars and halo, $\fj=1$, e.g. used by \citetalias{MMW98}, is
obtained if $s=0$.

\section{Results} \label{sec:results}

In this Section we present the results of our modelling technique and comparisons to
observations. In particular, we use a Monte-Carlo method to sample the distributions of
dark matter halo parameters (mass, concentration, spin), we generate a catalogue of model galaxies
and then we fit their scaling relations with power-laws. We start by comparing the predicted
scaling laws with the observations, then we investigate how their scatter is affected by the
model parameters and finally we compare the predicted correlation of the TF and MS
residuals with what is observed in SPARC.

Our aim here is not to find the best fitting parameters of the model and then
discuss their physical implications; instead, we just provide a proof-of-concept of
the fact that introducing a $\lambda-c$ correlation and an $\fj-\fM$ correlation has a
significant impact on the correlation of the TF vs. MS residuals, which can fully erase
them for a narrow range of parameters.
Thus, in what follows, we first fix the two free parameters of the model
($\rlc=-0.4$ and $s=0.4$) and explore its predictions, and later we show what is the effect
of varying these two parameters. A full fitting of the observations is left to future work,
with more parameters including a bulge component and a gas disc.

\begin{figure}
\includegraphics[width=0.5\textwidth]{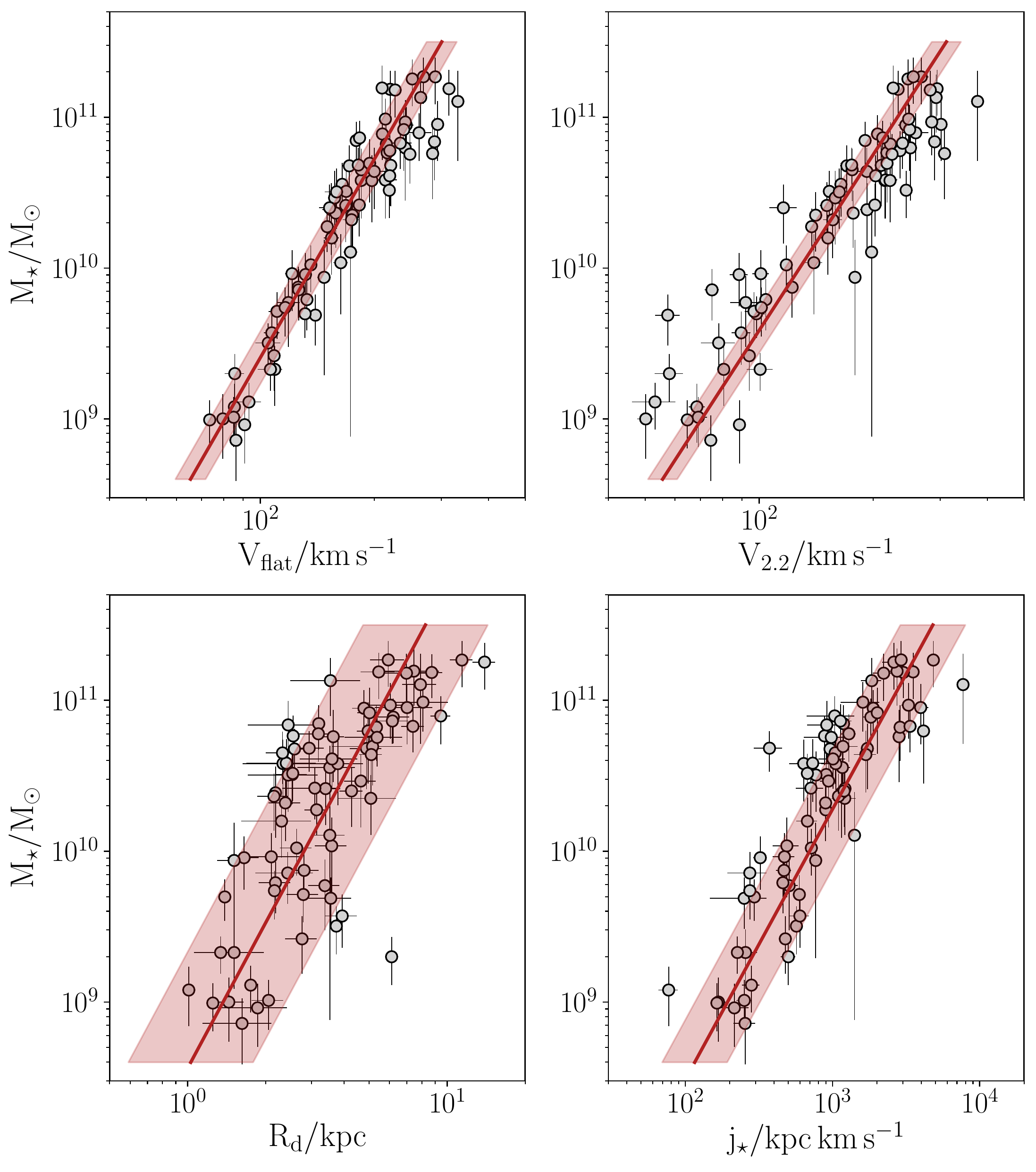}
\caption{Comparison of the predictions of the model with $\rlc=-0.4$ and $s=0.4$ (red lines)
        against the observations from the SPARC catalogue (grey circles).
        We adopt two different velocity definitions for the TF relation: $\vtwo$, the circular
        velocity at $2.2\Rd$, and $\vf$, which we compare with $\Vc(5\Rd)$ where our model
        rotation curves are approximately flat. The light red band shows the $1\sigma$
        intrinsic scatter of the model.
        }
\label{fig:SLcomp}
\end{figure}

\begin{figure*}
\includegraphics[width=\textwidth]{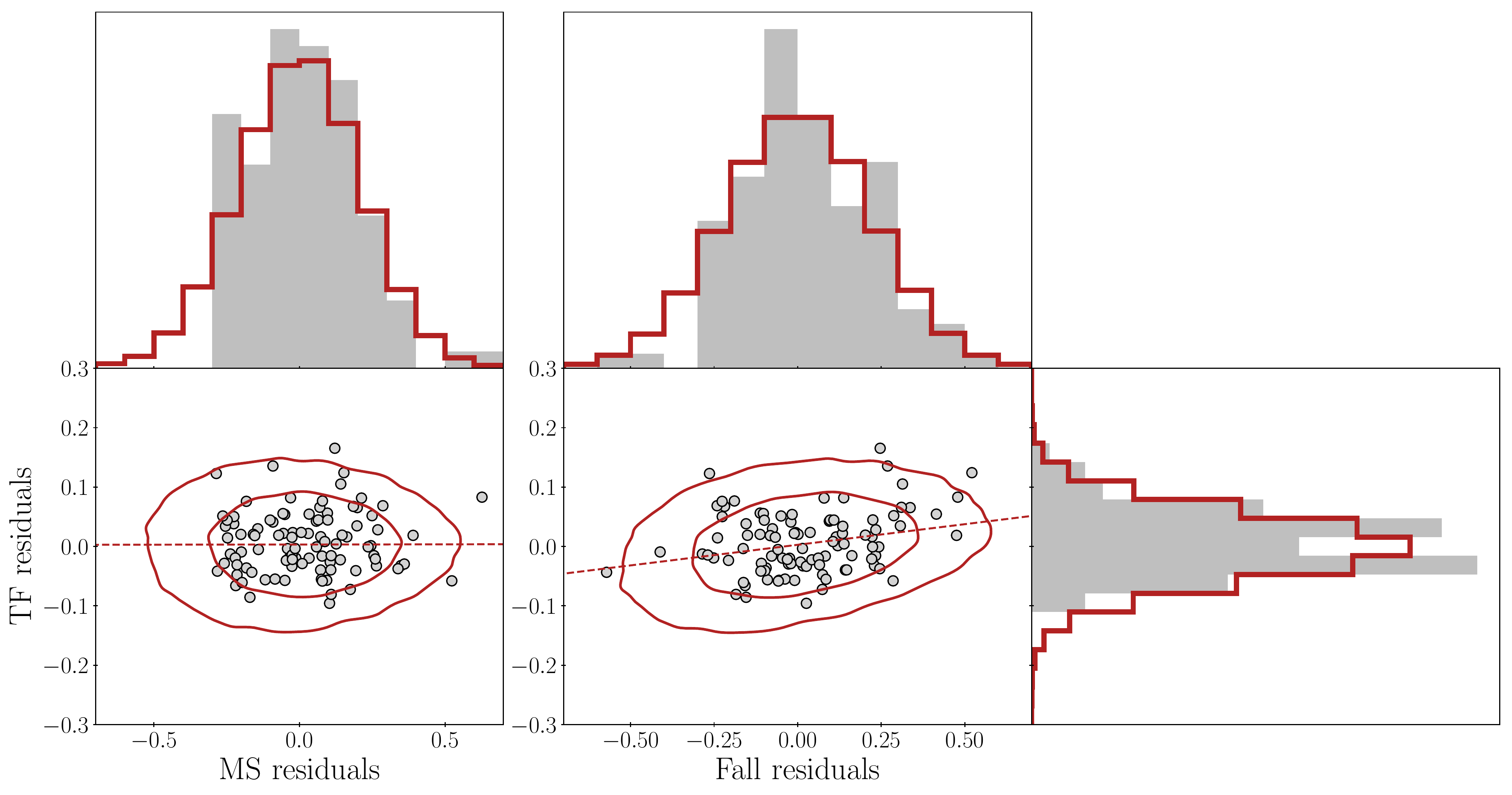}
\caption{Residuals of the Tully-Fisher vs. mass-size (left) and of the Tully-Fisher vs. Fall
         relation (right). Data from SPARC are shown as grey circles, while the red curves
         are the 1- and 2-$\sigma$ contours of the predicted galaxy distribution of the same
         model as in Fig.~\ref{fig:SLcomp}. The dashed line shows the slope of the correlation
         of the model. The histograms on top and on the right show the marginalised distributions
         of the respective residuals for the data (grey filled) and the model (red empty)
         respectively.
        }
\label{fig:residuals_comp}
\end{figure*}

\subsection{The Tully-Fisher, mass-size and Fall relations} \label{sec:sl_comp}

We now explore the predictions of the model, with fixed $\rlc=-0.4$ and $s=0.4$, on the
TF, MS and Fall relations. For the TF, we adopt two different radii to define the velocity
plotted in the TF diagram in comparison with the SPARC observations: $\RTF=2.2\Rd$ and
$\RTF=5\Rd$. While the former is a very typical choice, commonly used for TF studies
\citep[e.g.][]{CourteauRix99,Pizagno+07}, we compare the latter with observations of $\vf$,
the velocity in flat part of the rotation curves \citep[e.g.][]{Lelli+16a}, as at $5\Rd$
the circular velocities of our model galaxies are approximately constant.

Figure \ref{fig:SLcomp} shows the TF, MS and Fall relations for the SPARC sample (grey
circles) with gas fractions $\MHI/\Mstar\leq 1$ compared to the predictions of the model
(red lines). The agreement of this simple analytic model is remarkable and even the
intrinsic vertical scatter of the (stellar) TF of $\sim$0.05 dex in $\vf$ is almost
consistent with that estimated on the dataset ($\sim$0.04 dex, using the procedure
outlined in  \citealt{Lelli+19}). 

It is interesting to notice that the agreement of the model with the TF relation is quite good
for both velocity definitions ($\vtwo$ and $\vf$), meaning that the shape of the model's rotation
curves are to first order representative of those of real spirals. Also the observed sizes
and angular momenta of spirals are in a relatively good agreement
with those expected by our analytic model of an exponential disc in an NFW halo.
The predicted MS and Fall relations of the model are, however, possibly slightly
shallower than what it is observed. This might be related to the fact that we do not have bulges
in our model: at fixed $\Mstar$, the presence of a bulge would make a galaxy be more compact and
have a lower specific angular momentum \citep[e.g.][]{RF12}.

The model predictions for the scaling laws are basically straight lines and this is mostly due
to two facts. The TF is straight because we employ a power-law stellar-to-halo mass relation
(thus monotonic in $\fM$), which is suggested by the rotation curve analysis of
\citet[][]{PFM19}
and provides a good description of the disc galaxy distribution\footnote{
It is important to specify that this is valid only for discs, as it is well known that the
stellar-to-halo mass relation has a different shape for different galaxy types
\citep[e.g.][]{Dutton+10,Rodriguez-Puebla+15}.
} \citep{Posti+19b}.
Nonetheless, the MS and Fall
relations could still be non-linear since they strongly depend on the luminous and dark matter
distribution within galaxies. In fact, to get also the MS and Fall relations straight it is
also important to have $\log\fj\propto\log\fM$ as well \citep{Posti+18a}.

\subsection{The Tully-Fisher and mass-size scatters and their correlation}

While it is not completely new that a simple analytic model of the type presented in
Sect.~\ref{sec:model} is able to predict relatively well the general structure of disc galaxies,
we now move to a more detailed analysis of the residuals of the disc scaling laws, on which
models of this kind have had less successful comparisons with data \citep{CourteauRix99,
Dutton+07,Desmond+19}. We show that when allowing the halo concentration and spin to be
anti-correlated ($\rlc<0$) and the stellar angular momentum fraction to the stellar mass 
fraction to be correlated ($s>0$), the models can actually predict residuals on the scaling
laws in good agreement with what is observed.

In Figure~\ref{fig:residuals_comp} we show the residuals on the TF as a function of the residuals
on the MS and on the Fall relations. Here we define residuals as $\Delta X = \log\,X - 
\log\,X_{\rm fit}(\mstar)$ for $X=\vf,\Rd,\jstar$, where the fits to the scaling relations
are computed with the procedure described in \cite{Lelli+19}. The SPARC galaxies are shown as
(grey) points, while the distribution of model galaxies is represented by the two (red) contours
encompassing respectively 68$\%$ and 95$\%$ of the total population. The Figure also shows
the histograms of the marginalised distributions of the residuals of the three scaling laws
for the observed (grey) and model galaxies (red). From these histograms it is clear that the
scatter predicted by the model agrees very well with that of the SPARC sample, which perhaps
only has a slightly tighter Fall relation than expected (0.20 dex against 0.23 dex of the model).

\begin{figure}
\includegraphics[width=0.5\textwidth]{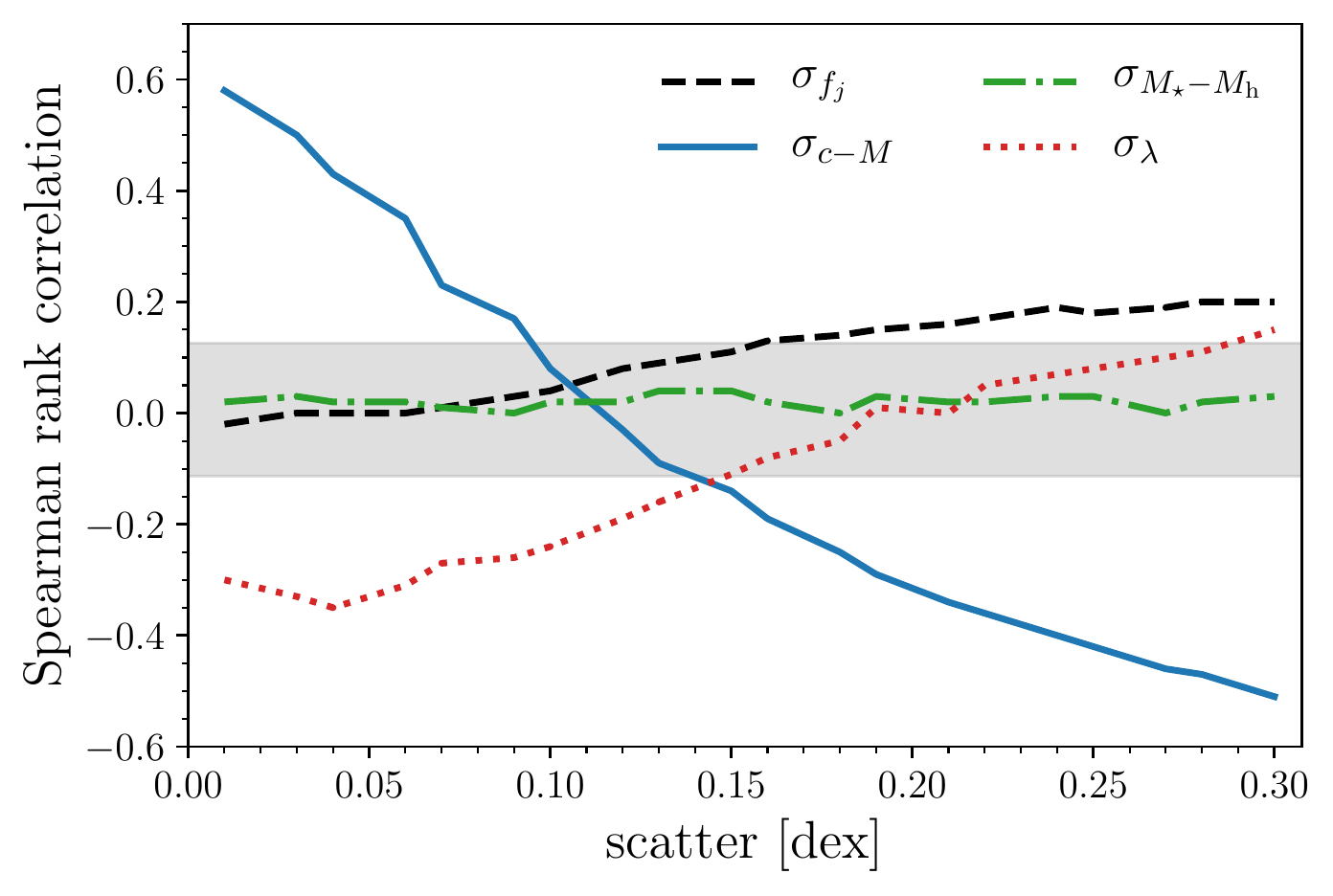}
\caption{Effect of varying the four intrinsic scatters of our model, 
         $\sigma_{M_\star-M_{\rm h}}, \sigma_{c-M}, \sigma_{\fj}, \sigma_\lambda$, on the
         Spearman correlation coefficient of the residuals of the TF vs MS. Each curve is
         computed varying only one scatter at a time, while leaving the others fixed to
         their fiducial values. The grey band shows the 1-$\sigma$ uncertainty (estimated
         with a bootstrap) of the correlation of the observed residuals of the TF vs. MS
         in SPARC.
        }
\label{fig:spear_allscatters}
\end{figure}

More importantly, the model presented here has residuals on the TF and MS that are not correlated,
as shown by the dashed red line in the bottom left panel of Fig.~\ref{fig:residuals_comp}. For
this model, the Spearman's rank correlation coefficient of the TF and MS residuals is negligible
($-0.03 \pm 0.02$, where the uncertainty has been estimated with a bootstrap technique).
This proves that simple, semi-empirical models where the sizes of discs are physically
linked to their angular momentum can be made compatible with current observational data on the
sizes and rotational velocities of discs.
Similarly, the model is also compatible with the shallow correlation that is observed between the
residuals of the TF and Fall relations. This is present also in the observed scaling laws simply
because the specific angular momentum of the discs is not independent on their rotation velocity.

\subsection{The effect of the model's intrinsic scatter on the TF vs. MS residuals}
\label{sec:scatter_effect}
We explore here what is the effect of the scatters of the various ingredients of the model on
the relation between the residuals of the TF and MS. In particular, we show what happens if
we vary the scatter of one particular ingredient of the model, while the others are fixed,
amongst: i)  the stellar-to-halo mass relation ($\sigma_{M_\star-M_{\rm h}}$), the halo
mass-concentration relation ($\sigma_{c-M}$), the retained fraction of angular momentum 
($\sigma_{\fj}$) and the halo spin parameter ($\sigma_\lambda$).
These scatters are, in fact, an important property of the model which directly determine both
the scatters of the observed scaling laws and their residuals -- e.g. a model with no scatter
would predict scaling laws with null scatter and thus no residuals. 

\begin{figure}
\includegraphics[width=0.5\textwidth]{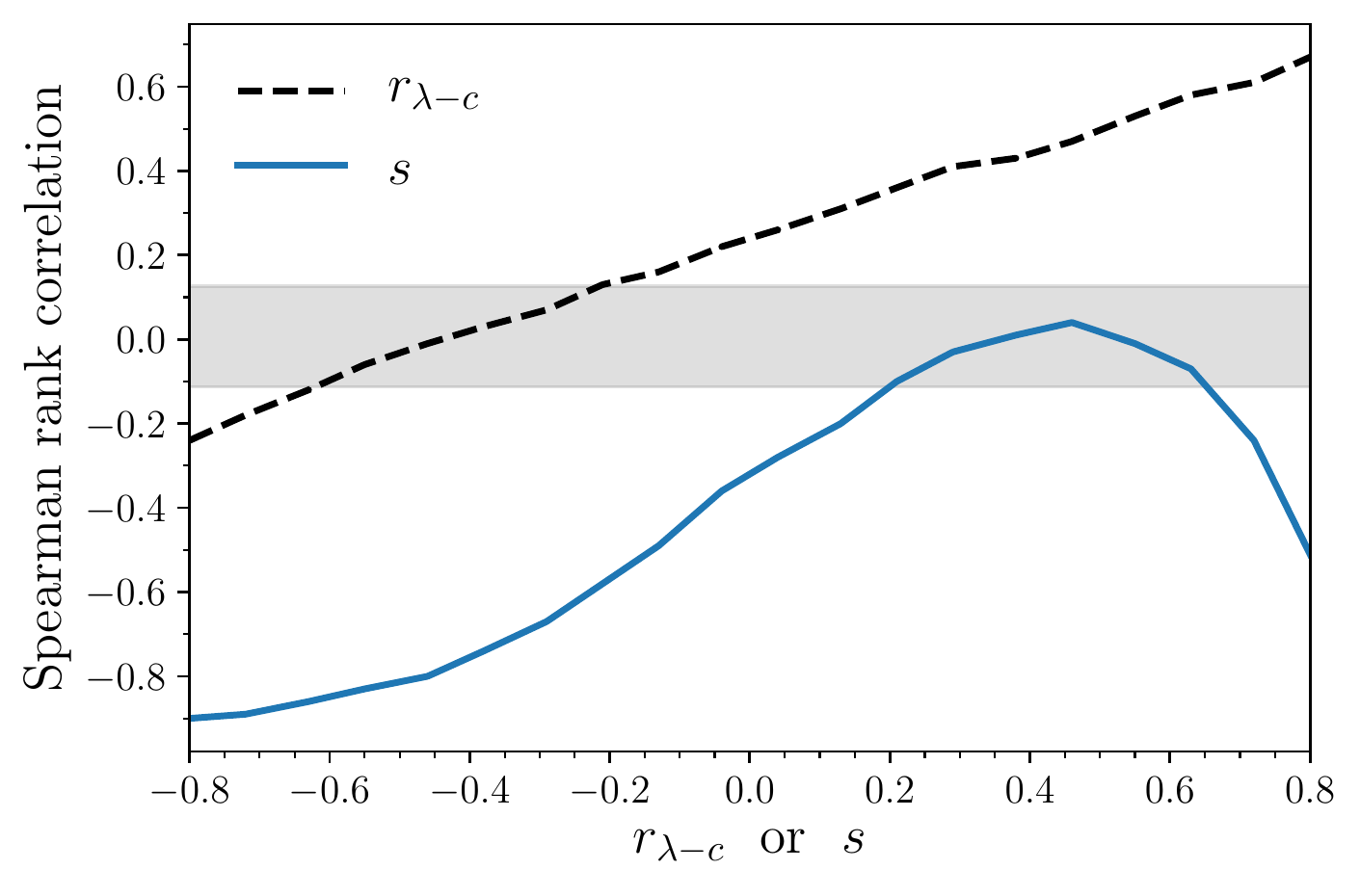}
\caption{Effect of varying the halo spin-concentration correlation ($\rlc$) and the
         power-law index of the relation between $\fj$ and $\fM$ ($s$) on the
         Spearman correlation coefficient of the residuals of the TF vs MS. The grey band
         is as in Fig.~\ref{fig:spear_allscatters}.
        }
\label{fig:spear_rlc_s}
\end{figure}

\begin{figure*}
\includegraphics[width=0.5\textwidth]{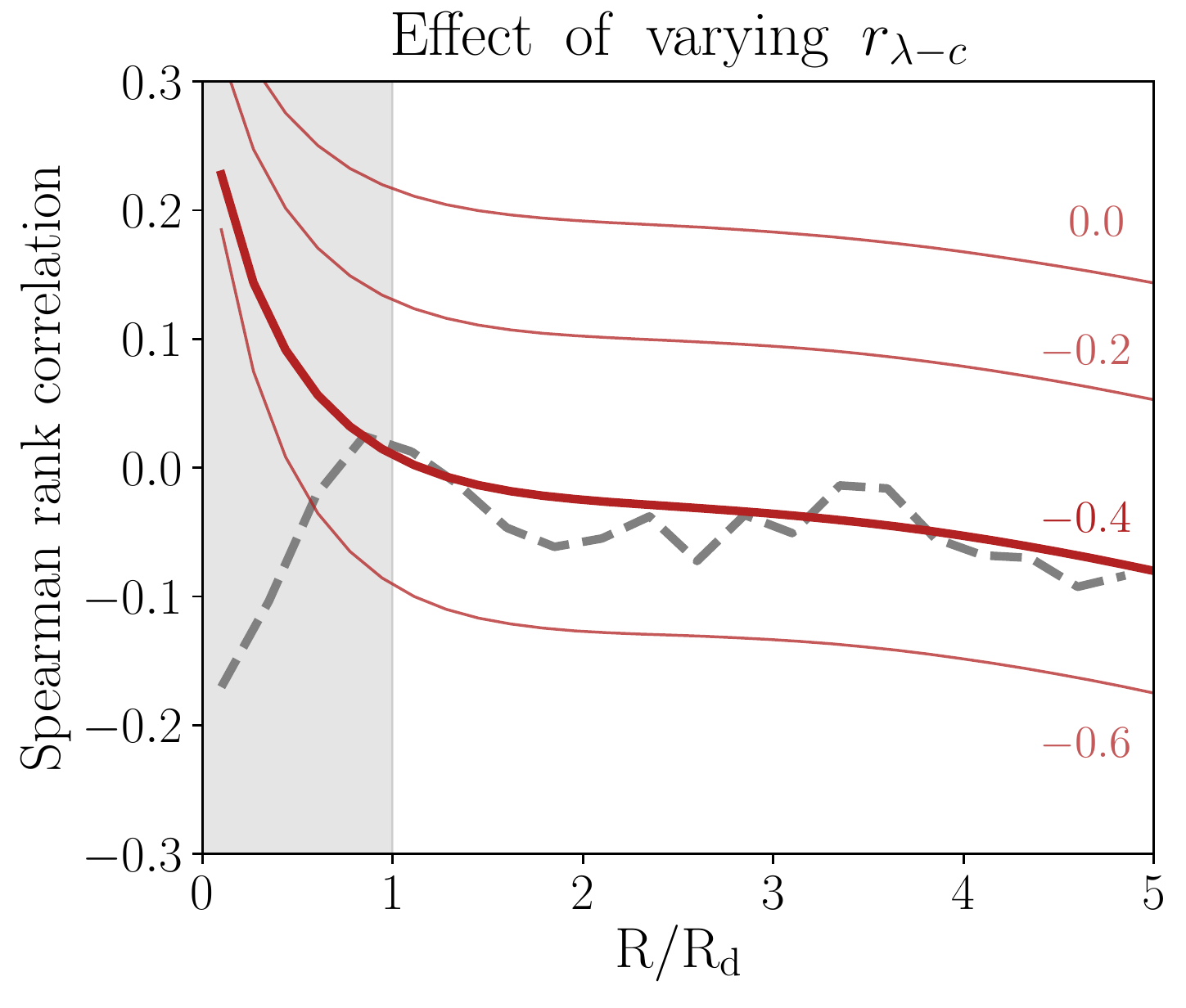}
\includegraphics[width=0.5\textwidth]{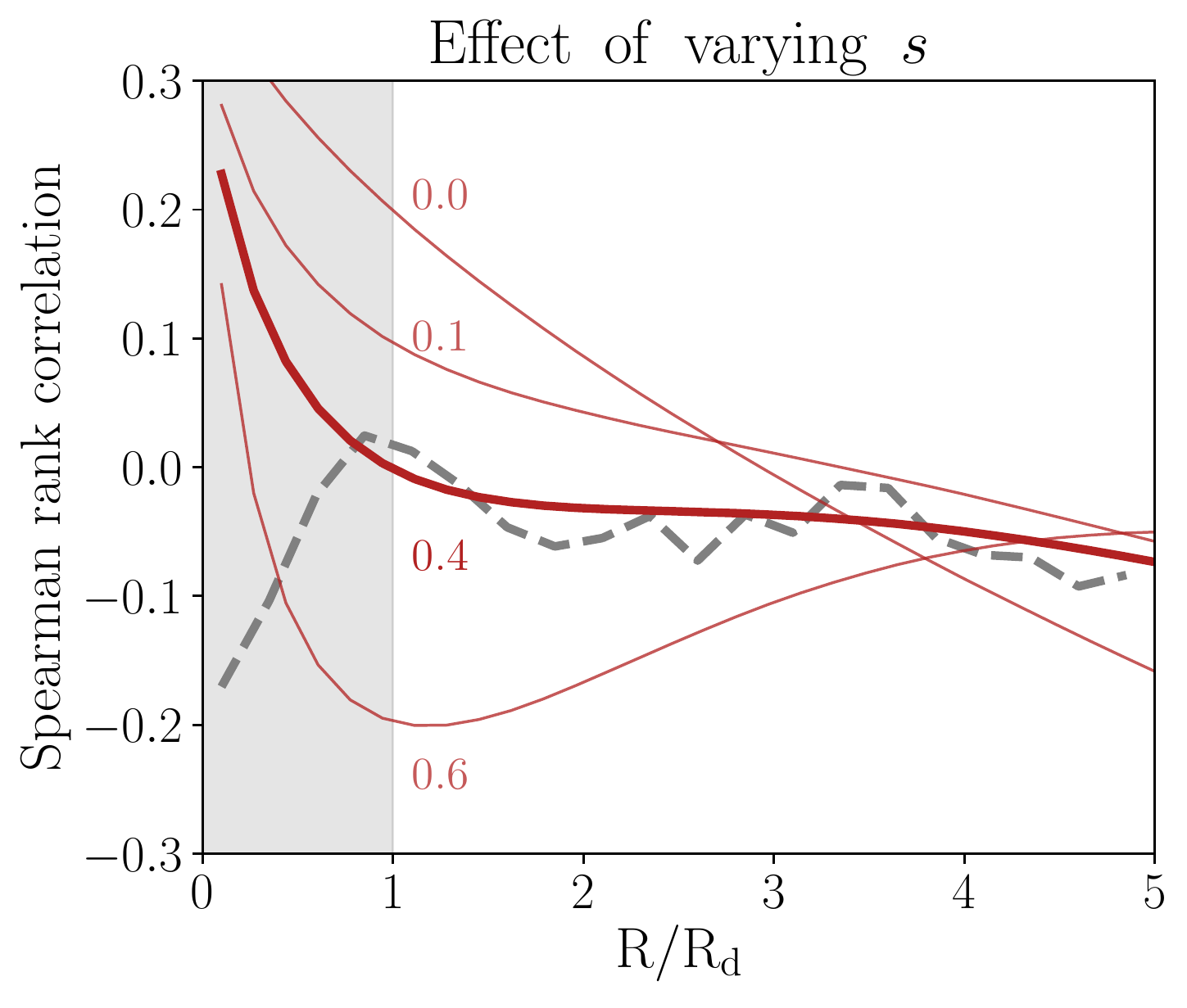}
\caption{Effect of changing the correlation coefficient between halo spin and concentration
         $\rlc$ (left) and the slope of the relation between the stellar angular momentum
         fraction and the stellar mass fraction $s$ (right) on the correlation of the TF
         and MS residuals as a function of radius where the TF velocity is measured.
         The grey dashed curve shows the data from the SPARC sample, while the different
         red curves are models with different values of such parameters. In both panels
         the thick red curve is our fiducial model (shown also in Figs.~\ref{fig:SLcomp}-
         \ref{fig:residuals_comp}) with $\rlc=-0.4$ and $s=0.4$; while in the left panel we
         fixed $s=0.4$ and in the right panel $\rlc=-0.4$. The grey area marks the region
         where the predictions of our model are not reliable, since we do not include a
         bulge component in our galaxies.
        }
\label{fig:residuals_radius}
\end{figure*}

However, even in a simple framework such as ours, the model scatters combine in a rather
non-trivial way, which makes it complicated to predict analytically what is the relation between
$(\sigma_{M_\star-M_{\rm h}}, \sigma_{c-M}, \sigma_{\fj}, \sigma_\lambda)$ and the resultant
residuals of the TF, MS and Fall relations. Thus we have numerically explored how the predictions
of the model vary as a function of the four scatters above while $\rlc$ and $s$ are fixed, in
order to build some intuition of their effect on the correlation between the residuals on the
TF and MS.
We show this in Figure~\ref{fig:spear_allscatters}. It is clear that while adopting a smaller
or larger scatter on the distribution of $\fj$ does not alter significantly the prediction of
the model, varying $\sigma_{c-M}$ and $\sigma_\lambda$, on the other hand, has a significant
effect on the correlation of the TF vs. MS residuals. We also find that the resulting Spearman
correlation coefficient does not significantly depend on the scatter of the $\Mstar-\Mh$
relation either: this is perhaps surprising since the $\Mstar-\Mh$ relation is the centrepiece
of the galaxy--halo connection, but it is somewhat reassuring since the value of its scatter
is still rather uncertain \citep[e.g.][]{WechslerTinker18}.

Let us now analyse in more detail some limiting cases. In what follows we discuss what happens
for model galaxies at a fixed $\Mstar$ and we probe their rotation curves at $\RTF=5\Rd$ to
define $\vf$.
\begin{itemize}
    \item[(i)] for $\sigma_{c-M}\simeq 0$, the range of concentrations of halos of different $\Mh$
               is very small as it is given only by the shallow slope of the $c-\Mh$ relation.
               At a fixed $\Mstar$, both high-$\lambda$ and high-$\Mh$ halos host galaxies
               with large $\Rd$ (because of Eq.\ref{eq:Rd}), thus they tend to have positive MS
               residuals. For a given $\Mstar$, large $\Rd$ also implies larger circular velocity,
               since the radius at which we probe the rotation curve, $\RTF$, is closer to the
               peak of the curve (the peak is also higher for high-$\Mh$ halos).
               These effects combine to produce a positive correlation of the TF vs. MS residuals.
    \item[(ii)] for a large $\sigma_{c-M}\,(\simeq 0.3$ dex), the range of concentrations
                spanned by halos in a given $\Mstar$ bin is instead large, such that variations
                in $\Vc$ are mainly caused by halos having different $c$. The larger $\sigma_{c-M}$,
                the more this effect is important over (i), so that high-$c$ halos tend to have
                positive TF residuals. Since $\lambda$ and $c$ are anti-correlated and since
                $\Rd$ scales with $\lambda$ and inversely with $c$ (via the factor $\xi$ in 
                Eq.~\ref{eq:Rd}), the TF vs. MS residuals become significantly anti-correlated.
    \item[(iii)] for $\sigma_\lambda\simeq 0$, the variation of the disc sizes is proportional
                 to $\fj$ and inversely proportional to $c$ (via the factor $\xi$ in
                 Eq.~\ref{eq:Rd}). Thus high-$c$ halos tend to have negative MS residuals. This,
                 together with the fact that high-$c$ halos have larger $\Vc$ and so positive
                 TF residuals, induces an anti-correlation between the TF vs. MS residuals.
    \item[(iv)] for a large $\sigma_\lambda\,(\simeq 0.3$ dex), the variations of $\Rd$ at a fixed
                $\Mstar$ are dominated by the variations in $\lambda$, such that
                high-$\lambda$ halos have positive MS residuals. In this case, $\Vc$ is
                significantly influenced by two factors: a) high-$c$ halos have larger $\Vc$ and
                b) high-$\lambda$ implies high-$\Rd$ and therefore high-$\Vc$, since the circular
                velocity in the TF is probed at a radius closer to its peak. While a) tends to
                induce an anti-correlation of the TF vs. MS residuals because of the $\lambda-c$
                anti-correlation, if $\sigma_\lambda$ is large enough b) becomes increasingly
                more important and tends to positively correlate the TF vs. MS residuals.
\end{itemize}

In any case, while the correlation of the TF and MS residuals depends significantly on 
the scatter of the halo concentration-mass relation, most numerical studies agree that the
plausible range for $\sigma_{c-M}$ is between 0.1 and 0.15 dex
\citep[e.g.][]{DuttonMaccio14,DiemerKravtsov15}. We emphasise that, the values of $\rlc$ and $s$
that give the best match to the observations will also slightly depend on the adopted value for
$\sigma_{c-M}$.

\subsection{The effect of varying the model parameters on the residuals across the rotation curves}

We now explore what is the effect of the two key parameters of the model, the spin-concentration
correlation $\rlc$ and the stellar fraction-angular momentum fraction correlation $s$. Since we
have full rotation curves both in the data and in the model, for completeness we show what is
the effect of varying $\rlc$ or $s$, while fixing the other, on the correlation of the TF and
MS residuals across the rotation curve \citep[similarly to][]{Desmond+19}. In particular,
we consider the rotation curve as a function of radius $R$, we fit the $\Vc(R)-\mstar$ relation
at that radius and we define the TF residuals as a function of radius as $\Delta \Vc(R) =
\log\,\Vc(R) - \log\,V_{\rm c,fit}(R|\mstar)$, where $\log\,V_{\rm c,fit}(R|\mstar)$ is
the velocity from the fit at a given radius and stellar mass. We do this for both the SPARC
data and our models.

First we show in Figure~\ref{fig:spear_rlc_s} the behaviour of the Spearman correlation
coefficient of the TF and MS residuals as a function of $\rlc$ and $s$, when the TF is evaluated
at $\RTF=5\Rd$. While the correlation coefficient monotonically increases for increasing $\rlc$;
it appears to be always negative for all values of $s$, with a clear maximum in the range
$0.2 \lesssim s \lesssim 0.6$. In this range of $s$, the values of the Spearman correlation
coefficient are compatible with what is observed in SPARC.

Figure~\ref{fig:residuals_radius} shows the comparison of the correlation coefficient
of the TF and MS residuals as a function of the radius where the TF is evaluated for the data
(grey dashed) and for our fiducial model (thick red). 
The model agrees very well with what is observed across the wide range
probed by rotation curves in the outskirts of spirals: both the model and the data have negligible
correlation of the TF and MS residuals for $R\gtrsim\Rd$. On the other hand, the model is not
reliable in the innermost regions ($R\lesssim \Rd$) since we do not include a bulge component.
In fact, in this simple model the $\Vc$ in the inner regions is typically still dominated by dark
matter; thus, at a fixed mass, the $\Vc$ of a galaxy with a larger $\Rd$ will be larger because
it is evaluated at a larger physical radius and $\Vdm$ rises close to the centre.

\cite{Desmond+19} already noted the fact that having basically uncorrelated residuals across the
rotation curve corresponds, in this framework, to an anti-correlation between the residuals on
the halo concentration $\Delta\,c$ and the disc scale length $\Delta\,\Rd$ at fixed stellar
mass ($\Delta c \simeq -0.5\,\Delta\Rd$ in the case of our model). 
However, while \cite{Desmond+19} imposes this correlation a posteriori to explain the
observed residuals, here it follows naturally from the correlations of parameters of the theory,
which may have well-defined physical origins: $\lambda$ and $c$ are correlated since halos
that have assembled later, and are therefore less concentrated, spin faster \citep[e.g.][see
also \citealt{Bett+07}]{Johnson+19}; $\fj$ and $\fM$ are correlated since star formation in
discs proceeds inside-out, collapsing material at progressively larger $j$ \citep[so-called
biased collapse, e.g.][]{DuttonvdBosch12,RF12,Kassin+12}.

The different thin red curves in Fig.~\ref{fig:residuals_radius} show the effect of varying
$\rlc$ (left panel) and $s$ (right panel) on the correlation of the TF vs. MS residuals as a
function of radius.
At fixed $\Mstar$, high-$\lambda$ halos tend to have positive MS residuals since they host
high-$\Rd$ galaxies. At the same time, high-$\Rd$ implies also positive TF residuals, since
the TF is probed in the rising part of the rotation curve, which leads to positive Spearman
coefficients. This can be significantly counteracted with a $\lambda-c$ anti-correlation:
in fact, if at fixed $\Mstar$ high-$\lambda$ halos have low-$c$, then their circular velocity
has a lower peak and this can lead to negative TF residuals, if the anti-correlation strength
$\rlc$ is strong enough. Both models with too high or too low $\rlc$ ($\gtrsim -0.2$ or
$\lesssim -0.6$ respectively) seem to be ruled out by the current data. The value of the sweet
spot, $\rlc\approx-0.4$, is instead compatible with state-of-the-art N-body simulations (see
Appendix \ref{app:bolshoi} and \citealt{Maccio+07}).

A similar behaviour, but slightly more complicated in the details, is observed if we vary
the power-law index ($s$) of the relation between $\fj$ and $\fM$. A model in which $\fj$
is constant, i.e. $s=0$, has a significant correlation of the TF and MS residuals that varies
strongly with radius, from positive to negative correlation. This effect is, again, mitigated
by an increasing value of $s$ that tends to make the correlation less prominent and more
constant as a function of radius, in better agreement with the SPARC data. We note that
some of the effects mentioned in Sect.~\ref{sec:scatter_effect} do depend in a rather
non-trivial way on radius (i.e. those related to $\Vc$ affecting the TF residuals) and it
is thus not surprising that their interplay will also depend non-trivially on radius,
leading to the behaviours presented in Fig.~\ref{fig:residuals_radius}.
However, of particular interest is the value of $s\sim 0.4$ at which we have the
sweet spot, since that is precisely the value that is required to match the
observations of the Fall relation\footnote{In fact, from
$\jstar\propto\fj\fM^{-2/3}\Mstar^{2/3}$ \citep[Eq. 5 in][]{Posti+18b},
if $\fj\propto\fM^s$, it follows that
$\jstar\propto\Mstar^{\frac{s}{2}+\frac{1}{3}}$, since roughly
$\fM\propto\Mstar^{1/2}$ \citep{Moster+13,PFM19}. The slope of the observed Fall relation
of spirals is therefore matched for about $s\approx 0.45$.} \citep[see][]{Posti+18b}.

\subsection{Limitations of our model}
Our results are useful to get a first order understanding of the importance of the
$\lambda-c$ and $\fj-\fM$ correlations in reproducing the observed disc scaling laws.
In this work, we showcased the effect of these two ingredients in a deliberately simple
galaxy formation model, with the purpose of isolating -- as much as possible -- the effect
that these new ingredients. Naturally, for this reason our model is far from being complete
and has a number of limitations that one should keep in mind.

For instance, \cite{Dutton+07} developed sophisticated semi-empirical models, in spirit
very similar to ours, to predict the TF and MS relations and the correlation of their
residuals.
With respect to what we have presented here their models neglect the possibility of a
$\lambda-c$ or of a $\fj-\fM$ correlation, however they do include a bulge component,
halo contraction and a prescription for the formation of stars out of a gaseous disc.
In their work they show what is the effect of all the ingredients that they include in
determining the shape of the TF and MS relations, as well as on the correlation of their
residuals, and they find that in principle they all play a role.
Our work should in fact be considered complementary to theirs, as we showed what is the
effect of two previously unexplored parameters ($\rlc$ and $s$) on the correlation of the
TF vs. MS residuals. Their effects should be dominant over those of the additional
ingredients that \cite{Dutton+07} included, at least for the galaxies we considered here:
all SPARC galaxies have relatively small bulges and we focussed on a radial range where
the bulge should anyway be sub-dominant ($R>\Rd$); in our analysis we excluded gas-dominated
discs and \cite{Desmond+19} already pointed out that halo contraction seems to have a minor
effect on the correlation of TF vs. MS residuals.
To make sure that the last point applies also to our models, we have run again our model
including also a prescription for halo contraction parametrised in the same way as in
\cite{Dutton+07}: we find that the effect it has on the correlation coefficient of the
TF vs. MS residuals is marginal with respect to the effect of $\rlc$ and $s$ in
Fig.~\ref{fig:spear_rlc_s}.

Our models can, and will in the future, be made much more predictive by adding some
of the additional ingredients mentioned above. For example, the absence of a bulge
component limits our predictive power in the inner regions of massive spirals and the
absence of a cold gas component limits our inference at the dwarf mass scales, where
galaxies are increasingly more gas-dominated \citep[e.g.][]{SPARC}.
In particular, the fact that the baryonic TF relation appears tighter than the stellar
TF relation might indicate that the cold gas mass should, for reasons yet to be understood,
tightly correlate with halo mass at a given stellar mass \citep[e.g.][]{Desmond17}.
Also, in our model stars are assumed to be on circular orbits, while in reality some asymmetric
drift is present in real galaxies and can in principle modify the stellar specific angular
momentum from that in Eq.~\eqref{eq:jstar}, especially for low-mass discs \citep[e.g.][]
{Posti+18b,ManceraPina+20}. 

\citet[][see also \citealt{FirmaniAvilaReese00}, \citealt{vandenBosch00}]{Dutton+07}
noticed that if star formation and surface density are related this impacts the scaling
laws, since a halo with a larger spin will form a larger disc, with lower surface density,
thus forming less stars. This effect, that is not considered in our model, induces at a
fixed $\Mh$ an anti-correlation of $\Mstar$ with $\lambda$ which, combined with a
$\lambda-c$ anti-correlation, makes galaxies residing in halos with different $\lambda$
to scatter approximately along the TF. While potentially important for understanding the
residuals of the TF, this effect is based on the idea that the disc total mass (gas+stars)
to halo mass relation is more fundamental than the stellar-to-halo mass 
relation and on a star formation law with a fixed density threshold. This might however
not be the case if, for instance, the link of disc mass to halo mass is actually determined
by the self-regulatory action of star formation, which primarily sets the stellar-to-halo
mass relation regardless of the specific form of the star formation law \citep[e.g.][]{Lilly+13}.
Therefore, the importance of the star formation law in setting the scatter of scaling
relations appears to be an interesting possibility, which however needs further scrutiny
including a complete treatment of feedback.

Recently, \cite{Jiang+19} used cosmological hydrodynamical simulations to study the
relation between the specific angular momentum of galaxies and dark matter halos and
found evidence for a weak correlation, due to a combination of complex phenomena
that lead to the formation of galaxies \citep[see also][]{Danovich+15}. 
These results are potentially very interesting since they revisit the physical basis of
the \citetalias{MMW98} study; however, it is yet to be demonstrated that they can reproduce
the observed Fall relation, since from their main result $\jstar/\jh\propto\lambda^{-1}$
it would follow that $\jstar\propto\Mh^{2/3}$ -- because $\lambda\propto\jh/\Mh^{2/3}$ --
which is not compatible with observations of the Fall relation, unless in the case of
a quasi-linear stellar-to-halo mass relation ($\Mstar\propto\Mh$) which is excluded
by the data \citep{Posti+19b}.
In any case, their results highlight that while the classical framework of \citetalias{MMW98}
is capable of representing the overall shape of the scaling laws, the physics it describes
is inevitably limited and its results should be taken with a grain of salt. The advantage
of the \citetalias{MMW98} framework is that it encapsulates all the complexity of galaxy
formation into a couple of simple parameters, $\fj$ and $\fM$, and it is successful since the
observed galaxy--halo connection is indeed overall simple \citep{Posti+19b}.

\section{Summary and Conclusions}

Galaxy disc scaling laws can extensively be used to provide powerful constraints to galaxy
formation models. For instance, the observed absence of correlations between the residuals
of the TF and MS relations has been claimed to pose a challenge to traditional analytic
models based on the assumption that disc sizes are regulated by halo angular momentum. 
In this contribution, we revisit this issue and we show that 
including correlations amongst some parameters of the galaxy formation model, which
have some physical grounds, can help in reproducing what is observed.
Our aim here to provide a proof-of-concept of the fact that the inclusion of previously
unexplored correlations of the theory's parameters has a significant effect in the prediction
of the disc scaling laws.
In summary, we find that:
\begin{itemize}
    \item if we allow the halo concentration to be anti-correlated to halo spin
             (as suggested by N-body simulations, e.g. \citealt{Maccio+07}) and the
             stellar-to-halo specific angular momentum fraction to be correlated to the
             stellar-to-halo mass fraction (as it is needed to reproduce the observed
             angular momenta of galaxies, e.g. \citealt{Posti+18b}); a simple
             semi-empirical model, where disc sizes follow from the disc angular
             momentum, can have correlations of TF-MS residuals and TF-Fall residuals
             as observed;
    \item the introduction of an anti-correlation between halo spin and concentration
             induces an anti-correlation between disc size and concentration, which in turn
             is needed to wash out the correlation between the residuals of the TF and MS
             relations. Thus, contrary to some recent claims, we were able to find a
             semi-empirical model based on the assumption that the halo angular momentum
             is related to that of the disc which correctly reproduces the scaling relations;
    \item the range of parameters $\rlc$ and $s$ (controlling the $\lambda-c$ and
             $\fj-\fM$ correlations) allowed by the observations is relatively tight.
             In particular, we find that the values of these parameters that provide the
             best representation of the observed galaxy distribution are interestingly
             compatible with the values expected by N-body simulations (Appendix
             \ref{app:bolshoi} and \citealt{Maccio+07}) and by previous works
             \citep[e.g.][]{Posti+18b}.
\end{itemize}

Despite the fact that the residuals of the galaxy scaling laws are an intrinsically noisy
observable, it is worthwhile modelling them since they carry unique constraints to galaxy
formation models. In order to surpass the current limitations given by the paucity of
high-quality data for dynamical studies of disc galaxies, it is, thus, imperative to
observationally measure the scaling laws on a much larger and, hopefully, complete sample
of spirals in the local Universe, albeit with similar quality, to properly be able to model
all of the facets of galaxy formation, which remains a difficult long-term challenge.

\begin{acknowledgements}
We thank the referee for a report that helped improving the quality and clarity of
this paper. We are grateful to Aaron Dutton for useful discussions.
LP acknowledges support from the Centre National d’Etudes Spatiales (CNES). 
BF and RI acknowledge funding from the Agence Nationale de la Recherche 
(ANR project ANR-18-CE31-0006 and ANR-19-CE31-0017) and from the European 
Research Council (ERC) under the European Union's Horizon 2020 research and 
innovation programme (grant agreement No. 834148). GP acknowledges support
by the Swiss National Science Foundation, grant PP00P2\textunderscore163824.
\end{acknowledgements}

\bibliographystyle{aa} 
\bibliography{refs}

\appendix

\section{$\lambda-c$ correlation in the Bolshoi-Planck simulation} \label{app:bolshoi}

\begin{figure}
\includegraphics[width=0.5\textwidth]{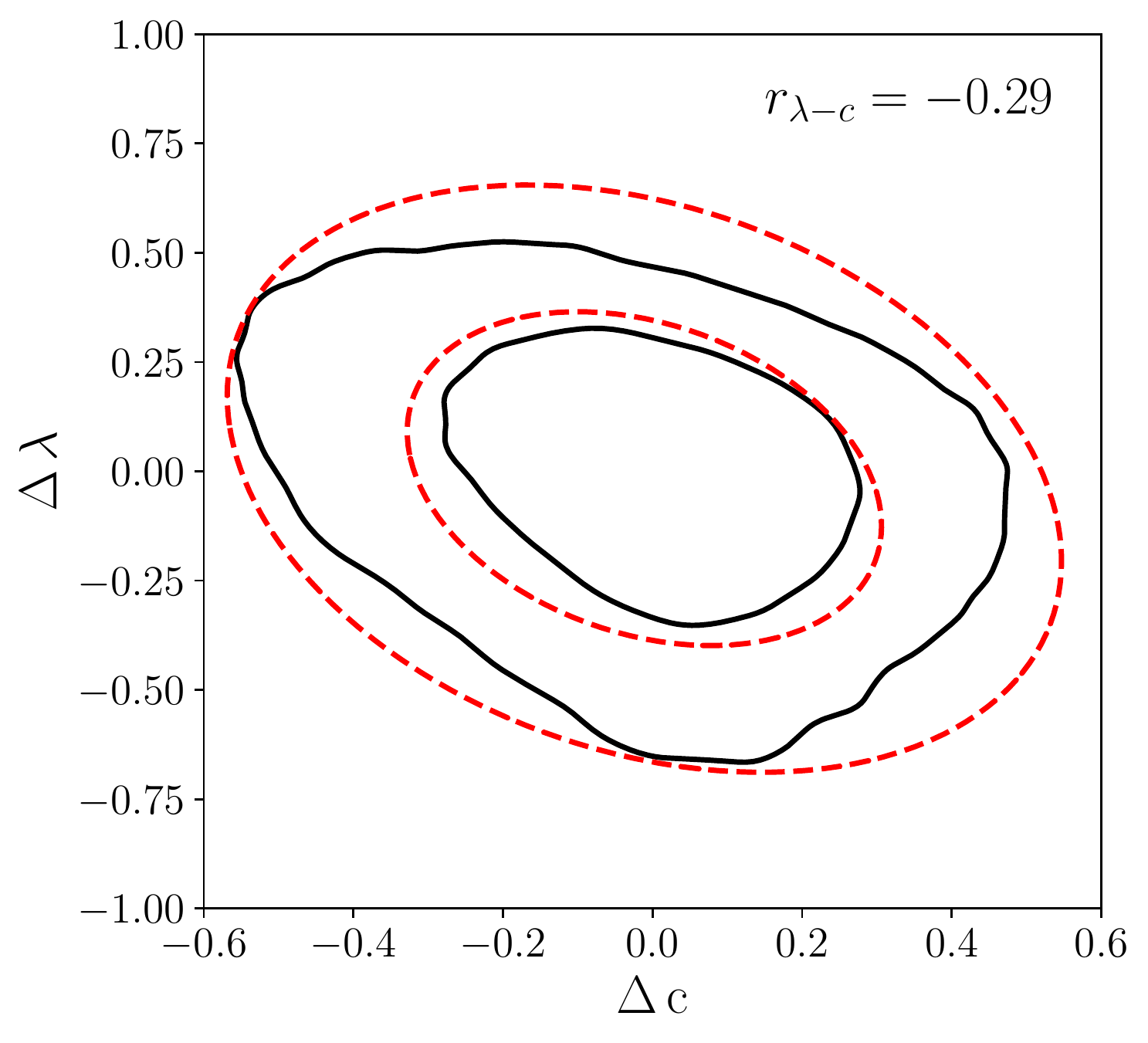}
\caption{Correlation between the $\lambda$ and $c$ residuals, at a fixed halo virial mass,
         for the $z=0$ halo population in the Bolshoi-Planck simulation, represented with
         the black solid contours (containing 68\%-95\% of the halo population). The red
         dashed contours are the 1- and 2-$\sigma$ contours of the 2D normal distribution
         defined with the covariance matrix calculated from the distribution of points in
         this plane.
        }
\label{fig:bolshoi}
\end{figure}

We used the publicly available catalogues of the Bolshoi-Planck simulation provided by
\cite{Rodriguez-Puebla+16} to have a simple estimate of the strength of the $\lambda-c$
correlation for halos in a dark matter only simulation with the standard Planck cosmology.
We considered the $z=0$ snapshot of the simulation where the halos where identified and
characterised with the \texttt{ROCKSTAR} software \citep{Behroozi+13b}.

We calculated the residuals at a fixed halo virial mass $\Mh$ of the $\lambda-\Mh$ and
$c-\Mh$ relations, where $\lambda$ is defined as in \cite{Bullock+01}. We show in
Figure~\ref{fig:bolshoi} the correlation of the $\Delta\,\lambda$ and $\Delta\,c$
residuals for the halos in the simulation with the black solid contours containing
68\%-95\% respectively of the halo population. We then calculate the covariance matrix
of the distribution of points in this diagram and we use this matrix to define the
2D normal distribution shown with the red dashed contours in Fig.~\ref{fig:bolshoi}.
This Gaussian has standard deviations of $\sigma_{\log\,c} \simeq 0.18$ dex and
$\sigma_{\log\,\lambda}\simeq 0.25$ dex \citep[consistent with previous estimates, 
e.g.][]{Maccio+07} and a correlation coefficient of $\rlc\simeq-0.29$.
Since our analysis is very simple, this number should just be used to have a rough idea
of what is the correlation coefficient that is expected for the halo population in
$\Lambda$CDM.

\end{document}